\documentclass{ws-p8-50x6-00}

\begin{document}

\title{The GENIUS Project - Background and technical studies}

\author{L.~Baudis, G.~Heusser, B.~Majorovits\thanks{talk presented by
    B\'ela Majorovits}, Y.~Ramachers,
  H.~Strecker, H.V.~Klapdor-Kleingrothaus} 

\address{Max Planck Institut f\"ur Kernphysik, Heidelberg, Germany}

%

\maketitle

\abstracts{The potential of GENIUS as a dark matter detector
  is discussed. A study was performed to demonstrate the good behaviour of the
  proposed detector design of naked HPGe-crystals in liquid
  nitrogen. The expected background components were simulated and are
  discussed in some detail.With the obtained background GENIUS could
  cover a large part of the favoured MSSM parameter-space.}


\section{Introduction}
In modern physics one of the biggest remaining challenges is the
question of the nature of dark matter. Enormous efforts are being
made to solve this puzzle. 
One of the most promising theoretical
candidates is the neutralino as the lightest supersymmetric particle
(LSP) (see, e.g.\cite{bedny,jung}). 

Neutralinos can be directly detected through elastic scattering off
nuclei in a low background detector.

The present experimental situation concerning direct detection of the
LSP is depicted in Fig.\ref{limits} (see \cite{yorck}). The area above the
solid lines 
represents the excluded parts of the M$_{WIMP}$-$\sigma_{0}$
parameter space from running experiments
(DAMA, Heidelberg-Moscow, CDMS)\cite{dama,ang2,cdms}. 
 
Also shown in the figure are the expected 
sensitivities of experiments presently under construction
(CDMS, CRESST, HDMS)\cite{cdms,cresst,HDMS}. The scatter plot
represents allowed 
solutions for the M$_{WIMP}$-$\sigma_0$ parameter space from MSSM
calculations\cite{bedny}. Obviously the
sensitivities of the running and forthcoming experiments are by far
too low to cover a great part of the favoured MSSM
parameter space. In order to reach this goal  
an improvement of the present best sensitivities by three to four
orders of magnitudes is required. A promising approach to this problem 
is 
the recent Heidelberg GENIUS proposal\cite{beyond,hirsch,klap,CERN}, which
suggests application of 100 kg of 'naked' natural germanium detectors for cold
dark matter search or 1 ton of enriched $^{76}$Ge for hot dark matter
search in liquid nitrogen in an underground setup.
In this way all materials are removed from the close vicinity of the detector.
Operating HPGe-crystals directly in liquid 
nitrogen has been discussed earlier\cite{heusser}. It already has been shown
that the detectors work well under these
circumstances\cite{klap,jochen}.
    

The use of liquid nitrogen (another system has been discussed
recently\cite{zdesenko})
has the following advantages:
\begin{tabbing}
\hspace*{0.5cm}\=- \=If the tank containing the liquid nitrogen is
made big enough, the LN$_{2}$ \\
\>\> could serve as a shielding against external radioactivity.\\
\>- The cooling efficiency would be optimal, since the crystals
  would \\
\>\> be in direct contact with the cooling medium. \\
\>- LN$_{2}$ can be produced with a high purity. \\
\end{tabbing}    
\vspace{-0.35cm}
The LN$_{2}$ would basically be the only material surrounding 
the crystals thus avoiding many dangerous background sources.
Under these circumstances the required background index of 0.01
counts/(kg keV y) could be reached for the region relevant for dark
matter search below 100 keV. In a first phase it is planned to
operate 40 natural germanium-crystals with $\sim$2.5 kg each,
resulting in an effective mass of $\sim$100 kg detector
material\cite{klap,NIM}. A suitable place for the experiment is the Gran
Sasso Underground Laboratory.


\section{Technical possibilities}
\begin{figure}[t]
\epsfxsize=15pc 
\hspace*{2.7cm}\epsfbox{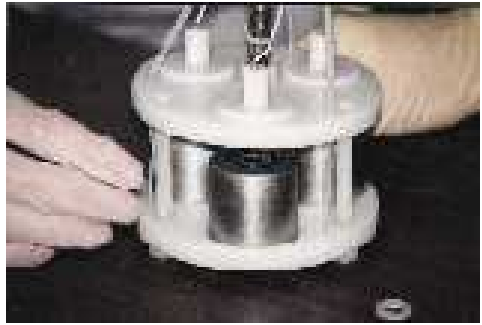} 
\caption{Setup of second study of Ge detectors operated inside liquid
  nitrogen. For details see text.\label{3detstudie}} 
\end{figure}

To prove beyond the first work\cite{klap,jochen} the technical feasibility of operating 'naked' HPGe crystals
directly in LN$_{2}$, two studies were performed. In
Fig.\ref{3detstudie} the three crystals of the second study are
shown. The cable length between crystal and FET (the first
preamplifier element) were 2m, 4m and 6m. Each crystal had a mass of
$\sim$300g. 

\subsection{Microphonics and interferences between detectors}
During extensive checks on the oscilloscope no signs of microphonics
could be detected. Also in an enforced test, when operating the detectors
directly after cooling them to their working temperature, despite the
strongly boiling 
LN$_{2}$ no signs, neither on the oscilloscope, nor in the recorded
spectrum, could be seen. Even when shaking the dewar the detectors did 
not show an alteration of their performance.
Furthermore there was no deterioration in the recorded background
spectra taken at the Gran Sasso underground laboratory with respect to 
the calibration measurements, which would be expected for
microphonics in the low energy region.

Moreover no signs of crosstalk between the detectors could be observed in the
recorded spectra.

\subsection{Electronic noise due to extreme cable lengths}
The obtained resolution for the detectors are listed in
Tab.\ref{tab:resolutions}. Comparing the results for the 'naked'
detectors with the performance of the 
conventionally operated 200g inner HPGe-detector of the HDMS
experiment\cite{HDMS} (see Tab. \ref{tab:resolutions}) an improvement
of the resolution has been 
obtained even for a cable length of 6m between crystal and FET. 
The threshold could be lowered to $\sim$2keV. The fact that the best
energy resolution
was reached for the detector with the longest cable length was due to
a change in the capacitance of the preamplifier which was only
performed for this 
detector, whereas for the other two the default setting was left
unchanged. 

\begin{table}[t]
\caption{Performance of the detectors\label{tab:resolutions}}
\begin{center}
\footnotesize
\begin{tabular}{
|c
|c
|r @{$\pm$} l 
|r @{$\pm$} l 
|c 
|}
\hline

Det \# 
& \parbox{1.5cm}{cable- length}
& \multicolumn{2}{c}{\parbox{2cm}{FWHM at 81keV [keV]}} \vline
& \multicolumn{2}{c}{\parbox{2cm}{FWHM at 356keV [keV]}} \vline
& \parbox{2cm}{threshold [keV]}
\\

\hline
\hline

1. Study
& / 
& 1.21 & 0.01
& 1.51 & 0.01
& ?
\\

\hline

99031 

& 4 m 
& 0.98&$^{0.06}_{0.19}$
& 1.12&0.02
& ?
\\


99032
& 2 m
& 0.82&0.05
& 1.01&0.02
& 1.85$\pm$0.20
\\


99033
& 6 m
& 0.68&0.05
& 1.04&0.02
& 1.85$\pm$0.20
\\

\hline

HDMS inner det.
& /
&1.0&0.05
&1.15&0.1
& 2.5 \\
\hline

\end{tabular}
\end{center}
\end{table}

\section{Expected background index}
To obtain an estimate of the awaited background, Monte Carlo
simulations with the code GEANT3.21\cite{GEANT} have been performed.

\subsection{External background sources}
The gamma flux in the Gran Sasso Underground
Laboratory\cite{borex_proposal}  was simulated 
leading to the conclusion that a
tank of 12m in diameter is needed to shield the detectors sufficiently 
from external gamma-radioactivity\cite{dipl}.

The neutron flux measured in the Gran Sasso\cite{belli} has to be taken into
account. This flux is initially reduced by the
polyethylene foam by 92\%. The through going neutrons will be mostly
thermalized and captured by the nitrogen through the reactions
$^{14}$N(n,p)$^{14}$C$^*$ and $^{14}$N(n,$\gamma$)$^{15}$N$^*$. The main 
contribution to the background from this component results from the
deexcitation of the nuclei. Considering the fact that the
thermalization takes part within the first 100 cm of the LN$_{2}$
shielding, the resulting gammas in the simulation where randomly
distributed in the first meter of the shielding. With an additional
boron-dotation of the polyethylene ($\sim$1000 kg) a count rate of
$\sim10^{-3}$ counts/(kg keV y) is expected.

The measured muon flux at the Gran Sasso Underground Laboratory is 1.1 
m$^{-2}$h$^{-1}$ with a mean energy of \=E$_{\mu}$=200GeV\cite{MACRO}.

The effect of muon-showers has been
simulated. By the use of a veto shield on top of the tank, reaching an
effectivity of 96 \%\cite{heusser2} this would yield
$\sim3\times10^{-3}$ counts/(kg keV y). This will be further reduced
through the anticoincidence between the 40 detectors in the setup.

The number of neutrons induced by muon showers is estimated 
to A$_n\sim3.2\times10^{-4}\frac{n}{gcm^{-2}}$ resulting in
$\sim2.5\times10^5$n/y in the whole tank\cite{berga}. The only 
produced long lived isotopes due to neutron-reactions are $^{14}$C
(T$_{1/2}$=5730y) and $^{13}$N (T$_{1/2}$=9.96m). The decay of the
$^{14}$C is negligible due to the long half life and the low decay
energy. The cross section for the production of $^{13}$N is by two
orders of magnitudes lower than the total cross section in the
relevant energy region, thus the contribution of these isotopes is
negligible. 

Other ways of producing long lived isotopes are $\mu$-capture by
the $^{14}$N nuclei ($\mu$+(Z,A)$\rightarrow$(Z-1,A)$^*$+$\nu_{\mu}
\rightarrow$ (Z-x,A-y)+n,p,$\alpha,\gamma$...) and inelastic muon
scattering ($\mu$+N $\rightarrow \mu^{'}$+X$^*$). The amount of produced
isotopes is less then 10$^4$ y$^{-1}$ in the tank and is therefore
negligible in 
comparison to the dominant background sources.

\subsection{Internal sources} 
Taking the purity level reached by the Borexino collaboration
for steel (5$\times10^{-9}$g/g $^{238}$U and
$^{232}$Th)\cite{borex_proposal}, a 
contribution of $\sim10^{-4}$ counts/(kg keV y) is expected from the
vessel.

The contribution of the LN$_{2}$ to the background mainly originates from
the nuclear decay chains U/Th, primordial $^{40}$K and $^{222}$Rn from
the surroundings. With the limits so far reached by the
Borexino-collaboration for purified water\cite{borexino} the decay
chains and $^{40}$K would contribute with $\sim4\times10^{-3}$
counts/(kg keV y) to the background between 0 keV and 100 keV. For the 
$^{222}$Rn contamination it is assumed that a purity of
100$\mu$Bq/m$^{3}$ LN$_{2}$ can be reached through an efficient isolation 
ensuring minimal diffusion of $^{222}$Rn into the tank leading to
$\sim5\times10^{-3}$ counts/(kg keV y) in the interesting region. Detailed
measurements of the purity of LN$_{2}$ are underway now, first results 
encourage the assumptions made.

The simulated holder system consisted of two 0.5 cm thick
high-molecular polyethylene plates being held by 5 rods of the same
material attached to the ceiling of the tank. Motivated by recent
results of the SNO collaboration\cite{SNO}, we assumed a radioactive
contamination of the polyethylene 100 times larger than the purity
reached by the Borexino collaboration for their liquid scintillator
material\cite{borexino}. With this we obtain
$\sim8\times10^{-4}$ counts/(kg keV y) in the relevant energy region.

To estimate the activation of the detector material through cosmic
radiation during production and transportation, the $\Sigma$
code\cite{bockholt} was used. The decays of  
the produced isotopes were simulated. In order to keep the
contamination low enough, it will be necessary to produce and
transport the crystals within 10 days if no additional shielding is
provided. The expected activities of these isotopes yield
$\sim10^{-2}$counts/(kg keV y).

\begin{figure}[t]
\epsfxsize=20pc 
\hspace*{1.75cm}\epsfbox{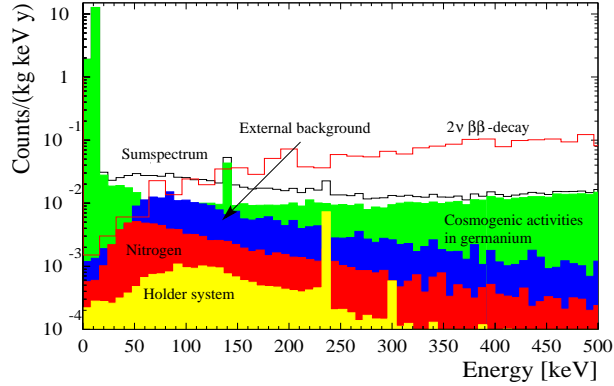} 
\caption{Resulting spectra of simulated components. Shown is also the
  contribution of the $2\nu\beta\beta$-decay, which can be well
  subtracted from the obtained spectrum. Due to the electron capture
  reaction of some cosmogenic nuclei and the following emitted
  K$\alpha$, the threshold of the detector will be $\sim$11
  keV.\label{all_sim}} 
\end{figure}

\section{GENIUS as a cold dark matter experiment}
The expected overall background is shown in 
Fig.\ref{all_sim}. It is evident that it should 
in principle be possible to reach a background index as low as a few
10$^{-2}$ counts/(kg keV y) between 10 keV and 100 keV. With this 
the GENIUS experiment could cover a major part of the favoured MSSM
M$_{WIMP}-\sigma_{0}$ parameter space (see also Fig.\ref{limits}).

\section{GENIUS as a hot dark matter experiment}
\begin{figure}[t]
\epsfxsize=15pc 
\epsfysize=18pc
\hspace*{-0.3cm}\epsfbox{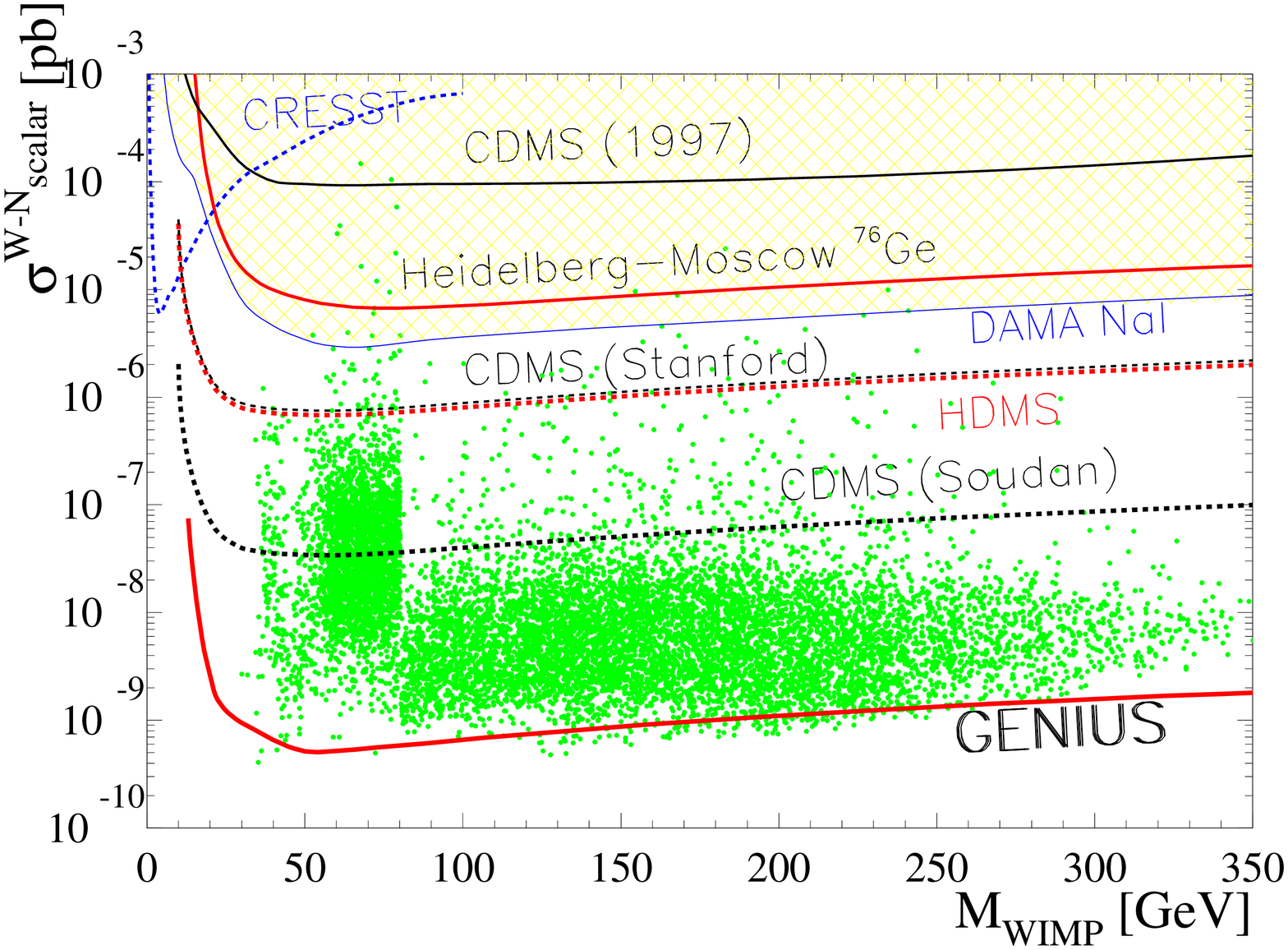} 
\epsfysize=18pc
\epsfxsize=14pc
\epsfbox{genalldeg_r05.epsi} 
\caption{Left: M$_{WIMP}-\sigma_0$ Limits and favoured region from
MSSM calculations (see text). Shown is also the evidence contour for a 
WIMP signal from the DAMA experiment. Solid lines correspond to
exclusion areas from running experiments, dashed lines to future
projects. Excluded is the area above the lines. Right: $\Delta$m$^2$-sin$^2$2$\theta$
exclusion plot. Excluded are the areas right of the curves. Shown is
also the potential of the GENIUS-experiment as a hot  
dark-matter detector. In case of degeneracy a 10 ton version of GENIUS 
could cover both, the small and large angle solution of the
solar neutrino problem (from$^{27}$).\label{limits}}
\end{figure}

The expected background for the region of the Q-value of the
0$\nu\beta\beta$-decay is $\sim5\times10^{-5}$counts/(kg keV
y)\cite{klap,jochen,dipl}. Using
1 ton of enriched $^{76}$Ge detectors would enable GENIUS to test
the effective Majorana neutrino mass 
down to $\sim$0.007 eV (68 \% C.L.) after ten years of
measurement\cite{klap}. Not only
would this allow to make a statement of the contribution of neutrinos to dark
matter, also conclusions on neutrino oscillations could be drawn, since
the 0$\nu\beta\beta$ decay observable $<$m$_{\nu}>$  
can be expressed in terms of oscillation parameters, if an assumption
of the ratio $\frac{m_1}{m_2}$ is made\cite{hirsch,klap}. For example in the
case of hierarchy
m$_1\ll$m$_2$, in a simplified model with two neutrino flavors, one
can write:
\begin{equation}
\Delta m_{21}^2 \sim m_2^2 =
\frac{4<m_{\nu}>^2}{(1-\sqrt{1-sin^22\theta})^2}. 
\end{equation}
For the case of degeneracy a 10 ton version of GENIUS with enriched
$^{76}$Ge could even cover the small angle MSW solution\cite{heinrich}(see
Fig.\ref{limits}).

\section{Conclusion}
As a preparational step of the GENIUS project,
a technical study has been made testing the performance of the new
detector design of naked HPGe-crystals operated directly in liquid
nitrogen. The attained properties are comparable or even better than
the ones of conventionally used HPGe-detectors. In particular we could 
not discover any signs of microphonics or interference. The
investigation of the 
awaited background components lead to the result that a background
index of a few 10$^{-2}$ counts/(kg keV y) can be reached in the
energy region below 100 keV relevant for the detection of WIMPs. This will
allow GENIUS to cover a major part of the favoured MSSM parameter space 
(see Fig.\ref{limits}) within 3 years of measurement with 100 kg of
natural HPGe detectors. Using a 1 ton version with $^{76}$Ge the effective
Majorana neutrino mass could be probed down to 0.01 eV within 1 year,
thus making GENIUS a powerfull tool in dark matter search and in the
search for other physics beyond the standard model.

\end{document}